\documentclass[11pt,preprint]{aastex}
\usepackage{chngpage}
\usepackage{natbib}
\usepackage{multirow}
\usepackage{longtable}
\usepackage{graphicx}

\shorttitle{Bimodal distributions in short GRBs}
\shortauthors{Yu et al. 2018}

\begin{document}

\title{Bimodal distribution of short gamma-ray bursts: evidence for two distinct types of short gamma-ray bursts}

\author{Y. B. Yu\altaffilmark{1, 2}, L. B. Li\altaffilmark{1, 2, 3}, B. Li\altaffilmark{1, 2}, J. J. Geng\altaffilmark{1, 2}, Y. F. Huang\altaffilmark{1, 2} }

\altaffiltext{1}{Department of Astronomy, Nanjing University, Nanjing 210023, China; hyf@nju.edu.cn}
\altaffiltext{2}{Key Laboratory of Modern Astronomy and Astrophysics (Nanjing University), Ministry of Education, China}
\altaffiltext{3}{School of Mathematics and Physics, Hebei University of Engineering, Handan 056005, China; lilongbiao@hebeu.edu.cn}

\begin{abstract}
GRB 170817A was confirmed to be associated with GW170817, which was produced
by a neutron star - neutron star merger. It indicates that at least
some short gamma-ray bursts come from binary neutron star mergers. Theoretically,
it is widely accepted that short gamma-ray bursts can be produced by
two distinctly different mechanisms, binary neutron star mergers and neutron
star - black hole mergers. These two kinds of bursts should be different
observationally due to their different trigger mechanisms. Motivated by this idea,
we collect a universal data set constituted of 51 short gamma-ray bursts observed by $Swift$/BAT,
among which 14 events have extended emission component.
We study the observational features of these 51 events statistically.
It is found that our samples consist of two distinct groups. They clearly show a bimodal
distribution when their peak photon fluxes at 15-150 keV band are plotted against the corresponding
fluences. Most interestingly, all the 14 short bursts with extended emission lie in a particular region of this plot.
When the fluences are plotted against the burst durations, short bursts with extended emission again
tend to concentrate in the long duration segment. These features strongly
indicate that short gamma-ray bursts really may come from two distinct types of progenitors.
We argue that those short gamma-ray bursts with extended emission come from the coalescence of
neutron stars, while the short gamma-ray bursts without extended emission come from
neutron star - black hole mergers.
\end{abstract}

\keywords{gamma rays bursts: general -- methods: statistical}

\section{Introduction}
\label{sect:intro}

Based on the short emission timescale, the energetics, and the wide variety of host
galaxy types (Berger 2014), it is generally believed that short gamma-ray bursts (GRBs) could
be connected with the merger of two compact objects due to the continuous loss of the orbital
energy and angular momentum through gravitational waves (\textbf{Paczynski 1986;}
Eichler et al. 1989; Narayan, ${\rm Paczy\acute{n}ski}$ \& Piran 1992; Gehrels et al. 2005;
Bogomazov, Lipunov \& Tutukov 2007).
There are actually two kinds of such compact binary mergers, i.e. black hole (BH) - neutron star (NS)
and NS-NS mergers (Janka, Ruffert \& Eberl 1998; Bloom, Sigurdsson \& Pols 1999;
Perna \& Belczynski 2002; Bethe, Brown \& Lee 2007; Tutukov \& Fedorova 2007).
Short GRBs produced by these two kinds of mergers may be slightly different (Dichiara et al. 2013; Lu et al. 2017).
Encouragingly, a NS-NS merger event (GW 170817) was recently directly detected by Advanced Laser Interferometer
Gravitational wave Observatory (LIGO) and Advanced Virgo and it was found to be accompanied by the short GRB 170817A (Abbott et al. 2017).
This NS-NS merger event occurs at a small distance of about 40 Mpc (Abbott et al. 2017), thus providing us
a valuable opportunity to study this kind of short GRBs (i.e., from NS-NS merger event). However,
no direct evidence for any short GRBs produced by NS-BH mergers has been observed till now.

The Burst Alert Telescope (BAT, Barthelmy et al. 2005a) on board the Neil Gehrels Swift Observatory (Gehrels et al. 2004) has detected more
than 100 short GRBs, above 70\% of which have also been detected by the X-Ray Telescope (XRT, Burrows et al. 2005).
We notice that a faction of short GRBs show an interesting feature, they are associated with extended emission (EE)
with timescale of about 100 s, which is much longer than the duration of the prompt emission (Barthelmy et al. 2005b;
Ioka et al. 2005; Norris \& Bonnell 2006). Although it is still unclear what is the origin of
such EE, there is no doubt that they are related to the activities of the central engines.

To explain the EE of short GRBs, various models have been suggested previously by many authors.
Considering that short GRBs with EE show no evidence for any association with a bright supernova,
Metzger et al. (2008) proposed that these events are related to proto-magnetars, which
are formed after the merger of binary NS systems, or are due to the accretion-induced collapse of white
dwarfs (WD), or even after the merger and collapse of WD-WD binaries. They argued that
the initial emission spike is powered by accretion onto the proto-magnetar from a transient
disk while the EE is produced by a relativistic wind that extracts the rotational energy of the
proto-magnetar. Bucciantini et al. (2011) further advanced this model by considering the effects
of surrounding material on the magnetar outflow. They even predicted a kind of phenomenon that 
the EE could be observable alone, without being associated with a short GRB.
Assuming that the observed energy release
during the EE is entirely due to the spin-down of the central magnetar, Gompertz et al. (2013)
found that the derived spin periods of the proto-magnetars, from the magnetic dipole spin-down model of
Zhang \& ${\rm M\acute{e}se\acute{a}ros}$ (2001), are in good agreement with that predicted for newly
born mangetars. Interestingly, recently Gibson et al. (2017) modified the magnetar propeller model to include
fallback accretion and used an MCMC method to fit the model to a sample of short GRBs exhibiting EE.
It is found that their model can cope with long, dipole plateaux and flare-like variability, indicating that fallback
accretion may play a pivotal role in explaining the features of EE light curves.

We believe that EE of short GRBs is crucial to reveal the physical origin of the central
engine as well as the radiation mechanism. Since short GRBs could naturally be divided into two groups
by considering whether they have extended emission or not, we conjecture that these two groups of
short GRBs may actually be produced by different kind of merger events.
More specifically, short GRBs with EE may be produced by binary NS mergers, while short GRBs without
EE are produced by BH-NS mergers. In this study,
we present a systematical analysis on a large set of short GRBs observed by $Swift$/BAT and
investigate their various parameters statistically to search for further evidence supporting
the classification.

The structure of our paper is as follows. In Section 2, we describe our sample. The
distribution of the fluence at 15-150 keV band and the peak photon flux at 15-150 keV band are
studied and the results are presented in Section 3. We summarize our
results in the final section with a brief discussion.

\section{Sample Selection}
\label{sect:obs}

We have collected the observational data of the short GRBs detected by $Swift$,
by using the 3rd $Swift$ BAT catalogue (Lien et al. 2016) and referring to the
compilation by Anand et al. (2017). To avoid the problem of having to consider
various biases caused by different instruments, we did not include bursts that were not detected
by the $Swift$-BAT. Our sample contains bursts only up to 2015, October,
when the BAT catalogue ends. In the catalogue, short GRBs with definite EE
were listed in their Table~3 by Lien et al. (2016). These authors also separately listed short GRBs
with possible EE in their Table~4. For these GRBs, the EE is either very weak, or quite
confusing due to bright X-ray sources in the field of view, or the GRBs themself are slightly
longer than 2s so that they are not typical short GRBs.
It is interesting to note that, the samples with EE in Lien et al's paper are from Norris
et al. (2010). In Norris et al's paper, the authors used the Bayesian Block methods to identify
the EE component. Firstly, they simulated bursts with EE to
calibrate the BAT threshold for EE detection and found a physical threshold
effect operates near $\rm R_{int} \sim few \times 10^{-3}$ ($\rm R_{int}$ is the ratio of average EE
intensity to initial pulse complex peak intensity), below which the EE
could not be identified. Based on this simulation, they finally got all the bursts with EE in the $Swift$/BAT sample.
In our study, we include all the short GRBs with definite EE in Lien et al's list into our sample.
Besides, in our data set, we add two other GRBs, 060614 and 080123, which are catalogued
as short GRBs with possible EE in Lien et al. (2016). GRB 060614 posed a great puzzle after
its detection. Being a long GRB ($\rm T_{90} \sim 100s$ in the BAT band), it is surprising that deep
searching for any underlying supernova gives null results (Gal-Yam et al. 2006; Fynbo et al. 2006; Della Valle et al. 2006).
It is quite unclear whether GRB 060614 is a collapsar-type event without supernova,
or a more energetic merger event (Gal-Yam et al. 2006).
To solve this enigmatic question, Zhang et al. (2007) conceived a ``pseudo''
burst that is about 8 times less energetic than GRB 060614 but with the same spectral
properties, based on the $\rm E_{p} \propto E_{iso}^{1/2}$ relation. They found that this
pseudo-burst would have been detected by BATSE as a marginal short-duration GRB, and would
have properties similar to GRB 050724 in the $Swift$ BAT and XRT bands.
Interestingly, although Anand et al. (2017) did not include GRB 060614 in their sample
due to its SN-less feature, both Kisaka et al. (2017) and L{\"u} et al.(2010)
put this burst in their sample as a GRB with definite EE. Considering all the above reasons,
we include this burst in our sample list.
As for GRB 081023, it is regarded as a event with definite EE by both Anand et al. (2017) and Kisaka et al. (2017).
Although GRB 090510A appeared in the sample of Anand et al. (2017), both Kagawa et al. (2015) and D'Avanzo et al. (2014) did not
treat it as a burst with EE. So we excluded this burst from our sample list.

As for the short GRBs without EE, we included all the bursts in the catalogue of Anand et al. (2017)
in our sample. In Anand et al.'s study, they did not include bursts with
confusing redshift information. These excluded bursts are GRBs 050813, 051105A, 051210, 060313, 070209, 070406,
091109B. In our statistics, the redshift information is not a key factor, so we include these GRBs in our sample.
Besides, two other bursts, GRBs 050416A and 080520, whose $\rm T_{90}$ are slightly longer than 2s, are also
excluded by Anand et al. (2017). However, in recognition of the fact that the duration
itself is likely to be an imperfect classification criterion for GRBs, Zhang et al. (2007)
resorted to the rest frame duration $\rm T_{90,z}$, which can be calculated
as $\rm T_{90,z} = T_{90}/(1+z)$. They argued that these bursts have rest durations shorter than 2s,
thus may be related to compact star mergers. Considering the fact that the rest frame durations of
GRBs 050416A and 080520 are both shorter than 2 seconds, we also include them in our sample.

In short, we use Lien et al. (2016) as the reference for EE bursts and our definite EE sample include
all Lien's EE bursts. We also include GRBs 060614 and 081023 as two weak but definite EE detections.
Other bursts classified as possible EE in Lien et al.'s Table 4 are not regarded as EE events
because the EE components are too uncertain.
As for the short GRBs without EE, our sample include all the bursts listed in the catalogue
of Anand et al. (2017). We additionally include GRBs 050416A, 050813, 051105A, 051210,
060313, 070209, 070406, 080520, 091109B in our study. We list all the bursts in our sample in Table 1.

\section{Bimodal Distribution}
\label{sect:nume}

Kagawa et al. (2015) classified the early X-ray
emission into two types after analyzing the time resolved spectrum of nine bright GRBs observed by $Swift$.
One is the EE with a rapid exponential decay during hundreds seconds since the SGRB trigger, the other is a
dim afterglow showing a power-law decay over $10^{4}$ s.
Lu et al. (2017) presented a comprehensive analysis of short GRBs observed by Fermi/GBM.
They identified two patterns of light curves among those GRBs. The duration and
$\rm E_{p}$ values of short GRBs observed with Fermi/GBM show a tentative bimodal feature.
In this section, we analyze our data set of the 51 short gamma-ray bursts observed by Swift/BAT.

In Figure 1, we plot the 15-150 keV fluence versus the peak photon flux (also in
the 15-150 keV band). The red points correspond to short GRBs without EE while the black points
are for the definite EE events. It can be clearly seen that the fluence at 15-150 keV band is
directly proportional to the peak photon flux at 15-150 keV band.
But most strikingly, the red points and the black points lie in two distinct regions.
The short GRBs with EE are in the right part of the panel while the short GRBs
without EE are at the left part of the panel.
They show a clear bimodal distribution, further proving that
they may be generated by two different types of progenitors.

In Figure 2, the 15-150 keV fluence versus the intrinsic duration are plotted for all our short GRB
samples. From this figure, we could see that there is a weak positive correlation between the observed fluence
and the duration. Again, the two kinds of short GRBs, i.e. those with EE and others without EE,
are in different regions of the plot. While short GRBs without EE distribute in a relative wide region in the panel,
short GRBs with EE are mainly concentrated on the top right of the figure. It further indicates
that short GRBs with EE may be intrinsically different.

In our Figures 1-2, the fluence was calculated over the $\rm T_{90}$ period, not including the extended emission portion.
For most of the bursts in our sample, the redshifts are available, so we can easily transfer the $\rm T_{90}$ in the
observer frame to that in the rest frame. As for the fluence, we lack their energy spectrum
information, therefore we can only show the fluence in the observer frame. It is interesting to
note that, though the fluence plotted in our figures are not in the rest frame, they are uniform.
In our plot, we include some GRBs with confusing redshift information, the rest frame $\rm T_{90}$ of which
can not be calculated, so the sample in Figure 2 is actually a little bit smaller than that in
Figure 1. Considering the fact that the Swift website does not provide error bars for the
z-dependent parameters, in our plot the properties of the bursts are shown without bars.

From Figures 1 and 2, we see that short GRBs with EE distribute differently from other short GRBs without EE.
They generally have relatively large fluences, large peak flux, and longer $T_{\rm 90}$ durations.
It means that the central engines of short GRBs with EE can last significantly longer and also be more powerful
than those short GRBs without EE.

According to Hotokezaka et al. (2012), in the case of a NS-NS merger, the total ejected rest mass can be as
high as $10^{-2}M_{\odot}$ ($\rm M_{\odot}$ is the mass of the Sun), although the exact value strongly
depends on the equation of state, the total mass as well as the mass ratio of the two neutron stars.
On the contrary, in the case of a BH-NS merger, the ejected mass is expected to be significantly less.
So, for a NS-NS merger, more energy can be released. It can naturally
lead to a short GRB that is brighter (with a higher peak flux) and more energetic. The burst will also
last for a longer duration. At a slightly later stage, when clumps of the ejected matter continuously fall back
toward the central compact star, a component of extended emission could be generated.
We thus suggest that short GRBs with EE are produced by NS-NS mergers, while short GRBs
without EE are produced by BH-NS mergers.


According to our theory, there should be an EE following GRB 170817A. However, no extended emission
was reported for this event. In Figure 1, GRB 170817A was also plotted. It lies in the boundary region
between GRBs with EE and those without EE, indicating that GRB 170817A may be a very special event. Actually,
it was argued that this burst should be produced by an off-axis jet (Evans et al. 2017; Kasliwal et al. 2017;
Margutti et al. 2017; Pian et al. 2017; Troja et al. 2017). The broadband X-ray to radio observations of the
afterglow of GRB 170817A can be well explained by such an off-axis jet model, with a viewing angle of
$\theta_{\rm obs} \sim 20^{\circ}-40^{\circ}$ and the inferred
kinetic energy being $\rm E_{k} \sim 10^{49-50}$ erg. So, GRB 170817A obviously
is not a typical short GRB. Then it is not strange that no obvious EE component was observed after the burst.

\begin{figure}
   \begin{center}
   \plotone{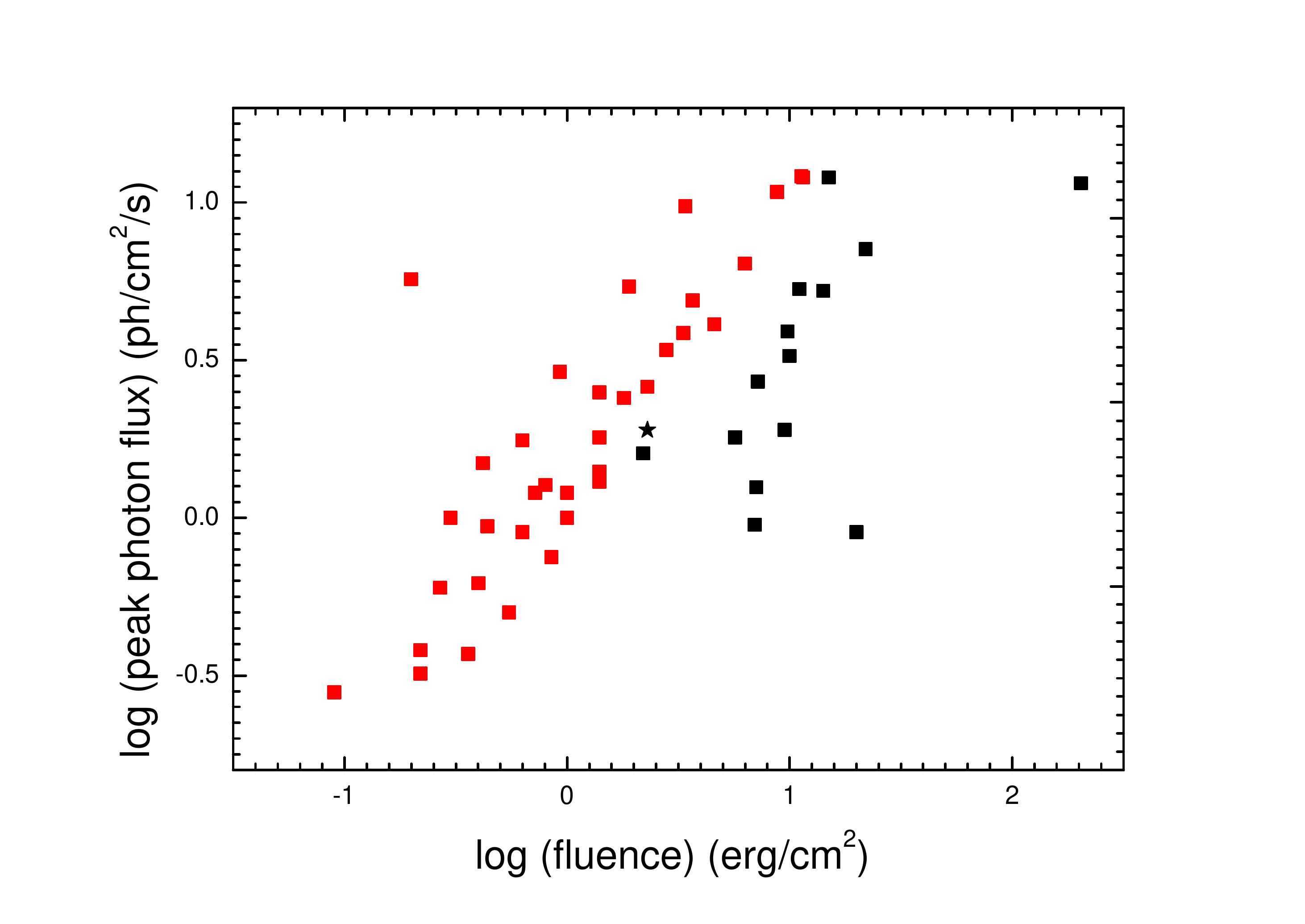}
   \caption{ The fluence and the peak photon flux observed at 15-150 keV for all the short GRBs in our data set.
   The black points are short GRBs with EE, while the red points
   are short GRBs without EE. The black star in the figure is the special event GRB 170817A.}
   \label{Fig:plot1}
   \end{center}
\end{figure}

\begin{figure}
   \begin{center}
   \plotone{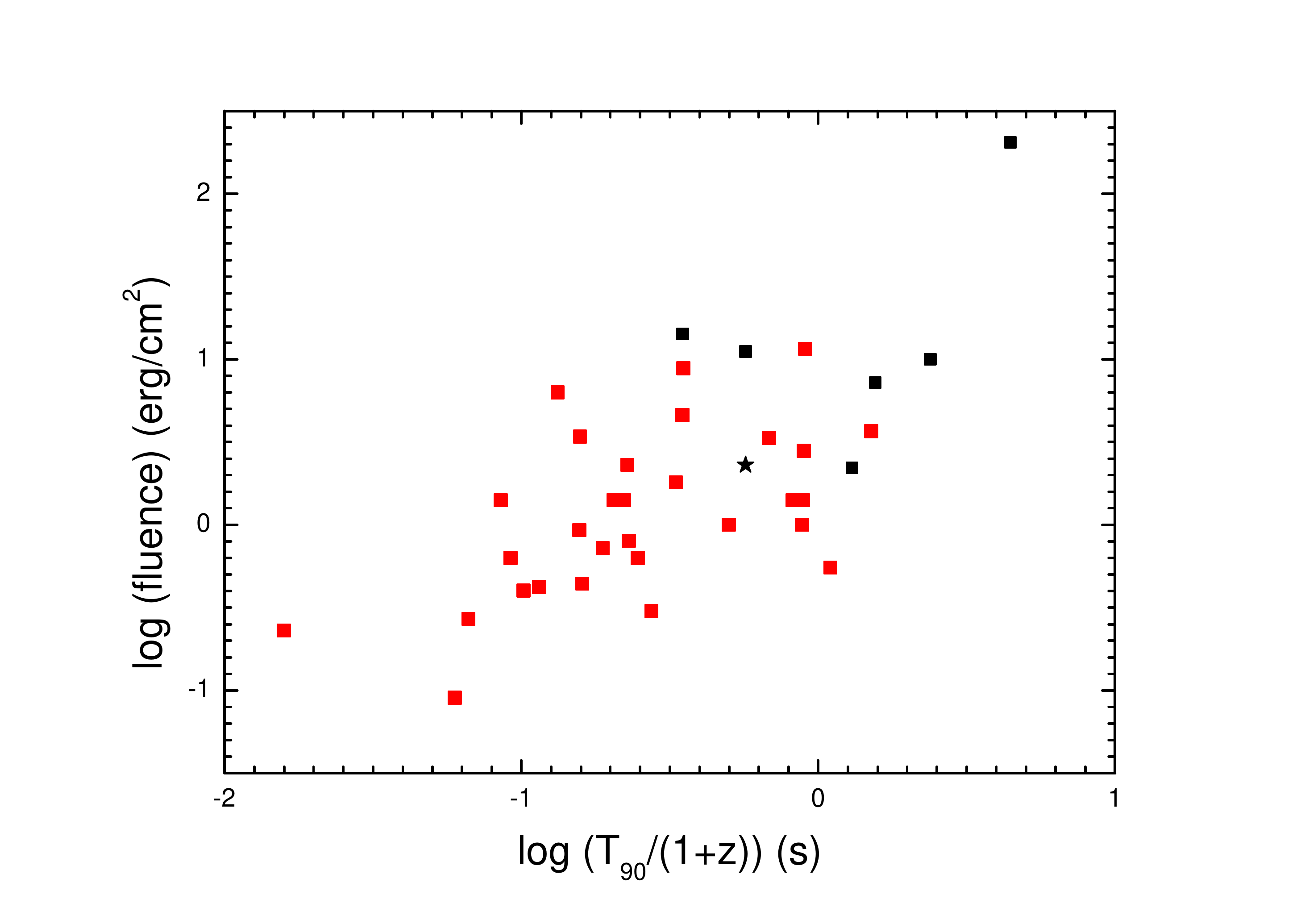}
   \caption{The rest frame duration and the fluence observed at 15-150 keV. The black points are short GRBs with EE while the red points
   are short GRBs without EE. The black star in the figure is the special event GRB 170817A.}
   \label{Fig:plot2}
   \end{center}
\end{figure}

\section{Discussion and Conclusions}
\label{sect:disc}

The most favorable progenitors of sGRBs are the coalescence of NS-NS binaries
(Paczynski 1986; Eichler et al. 1989) or NS-BH binaries (Paczynski 1991).
Motivated by the NS-NS merger origin of GW170817/GRB 170817A, we are interested in
searching for possible signatures supporting the existence of two different kinds of short GRBs,
i.e. those produced by NS-NS mergers and those produced by NS-BH mergers.
In this study, we have collected the key parameters for 51 short GRBs detected by $Swift$/BAT.
A fraction of these GRBs have obvious EE, while others do not.
It is found that when the fluence at 15-150 keV is plotted versus the corresponding peak photon flux,
the data points show an interesting bimodal distribution: short GRBs with EE distribute in a region
that is very different from other GRBs without EE.
On the duration vs. fluence plot, short GRBs with EE also show a tendency of lying in a
separate region. These facts strongly indicate that these two kinds of
GRBs (i.e. those with EE and those without EE) may be of different origin.
Since the EE component is usually connected with energy release due to accretion after the
prompt emission, we argue that short GRBs with EE should be produced by binary NS mergers,
while short GRBs without EE may be produced by BH-NS mergers.

It is interesting to note that a few other authors have also tried to identify the difference
between the two kinds of sGRBs produced by NS-NS mergers and NS-BH mergers.
For example, Dichiara et al. (2013) have tried to search for periodic and quasi-periodic signals
in the prompt emission of bright sGRBs. They argued that quasi-periodic jet precession could happen
in BH-NS mergers, but not in NS-NS mergers. They finally presented a zero result for periodic or
quasi-periodic signals for all their sGRBs, and concluded that BH-NS systems should be very rare
in the currently observed sGRBs.

In another study, Lu et al. (2017) also interestingly identified two kinds of sGRBs by analyzing the light
curves: Pattern I with single episode and Pattern II with multiple episodes. They found that
about 61\% of sGRBs belong to Pattern I and 39\% belong to Pattern II. They argued that Pattern I
events should be produced by NS-NS mergers and Pattern II bursts may be produced by another type of
compact binaries. Comparing the results of Lu et al. (2017) with those of Dichiara et al. (2013), we
see that the issue is still far from being resolved. In this study, we have tried to examine sGRBs
at a completely different aspect. We believe that the extended emission component would be a key
clue pointing to the nature of sGRBs. In the future, more studies on the extended emission may
give an answer to the physical origin of the central engines of different kinds of short GRBs.

\acknowledgments

\appendix
We thank the anonymous referee for helpful suggestions.
This work is supported by the National Natural Science Foundation of China (Grant Nos. 11873030 and 11873030),
the National Postdoctoral Program for Innovative Talents (Grant No. BX 2017M620199),
and the Strategic Priority Research Program ``Multi-wave band gravitational wave
Universe'' (Grant No. XDB23040000) of the Chinese Academy of Sciences.

\begin{deluxetable}{cccccc}
\tabletypesize{\scriptsize}
\tablewidth{0pt}
\tablecaption{List of samples.\label{TABLE:Fit1}}

\tablehead{
        \colhead{GRB name} &
        \colhead{Redshift} &
        \colhead{$\rm T_{90}$} &
        \colhead{$\rm peak~photon~flux (15-150 keV)$} &
        \colhead{$\rm Fluence (15-150 keV)$}  &
        \colhead{EE} \\
        \colhead{} & \colhead{} & \colhead{(s)} & \colhead{$(ph/cm^{2}/s)$} & \colhead{$(10^{-7}erg/cm^{2})$} & \colhead{(Yes or No)} }
\startdata
050416A            & 0.6535           & 2.5           & 4.88     & 3.67     & N \\
050509B       & 0.225           & 0.073           & 0.28     & 0.09     & N        \\
050724       & 0.2576            & 3.01      & 3.26  & 9.98  & Y  \\
050813     & 1.8            & 0.45            & 0.94      & 0.44    & N     \\
051105A               &             & 0.093            & 0.32      & 0.22 & N       \\
051210            &            & 1.3           & 0.75     & 0.85     & N \\
051221A            & 0.547           & 1.4           & 12     & 11.5     & N \\
051227            &            &            & 0.95     & 6.99     & Y \\
060313            &            & 0.74           & 12.1     & 11.3     & N \\
060502B           & 0.287           & 0.131           & 0.62     & 0.4     & N \\
060614            & 0.1254           & 4.997           & 11.5     & 204     & Y \\
060801            & 1.131           & 0.49           & 1.27     & 0.8     & N \\
050416A            & 0.6535           & 2.5           & 4.88     & 3.67     & N \\
061006            & 0.4377           & 0.50           & 5.24     & 14.2     & Y \\
061201            & 0.111           & 0.76           & 3.86     & 3.34     & N \\
061210            & 0.4095           & 0.80           & 5.31     & 11.1     & Y \\
061217            & 0.827           & 0.21           & 1.49     & 0.42     & N \\
070209            &            & 0.09           & 0.38     & 0.22     & N \\
070406            &            & 1.2           & 0.37     & 0.36     & N \\
070429B            & 0.904           & 0.47           & 1.76     & 0.63     & N \\
070714B            & 0.9225           & 3.0           & 2.7     & 7.2     & Y \\
070724A            & 0.457           & 0.4           & 1     & 0.3     & N \\
070729           & 0.8           & 0.9           & 1     & 1     & N \\
070809            & 0.473           & 1.3           & 1.2     & 1     & N \\
071227            & 0.394           & 1.81           & 1.6     & 2.2     & Y \\
080123            &            &            & 1.8     & 5.7     & Y \\
080503            &            &            & 0.9     & 20     & Y \\
080520            & 1.545           & 2.8           & 0.5     & 0.55     & N \\
080905A            & 0.1218           & 1           & 1.3     & 1.4     & N \\
090426            & 2.609           & 1.2           & 2.4     & 1.8     & N \\
090510            & 0.903           & 0.3           & 9.7     & 3.4     & N \\
090515            &            & 0.036           & 5.7     & 0.2     & N \\
090531B            &            &            & 1.25     & 7.1     & Y \\
090715A            &            &            & 3.9     & 9.8     & Y \\
090916            &            &            & 1.9     & 9.5     & Y \\
091109B           &            & 0.3           & 5.4     & 1.9     & N \\
100117A            & 0.915           & 0.3           & 2.9     & 0.93     & N \\
100206A            & 0.407           & 0.12           & 1.4     & 1.4     & N \\
100625A            & 0.452           & 0.33           & 2.6     & 2.3     & N \\
101219A            & 0.718           & 0.6           & 4.1     & 4.6     & N \\
111117A            & 1.3           & 0.47           & 1.35     & 1.4     & N \\
111121A            &            &            & 7.1     & 22     & Y \\
120804A           & 1.3           & 0.81           & 10.8     & 8.8     & N \\
130603B            & 0.356           & 0.18           & 6.4     & 6.3     & N \\
131004A            & 0.717           & 1.54           & 3.4     & 2.8     & N \\
140622A            & 0.959           & 0.13           & 0.6     & 0.27     & N \\
140903A            & 0.351           & 0.3           & 2.5     & 1.4     & N \\
141212A            & 0.596           & 0.3           & 1.2     & 0.72     & N \\
150101B            & 0.134           & 0.018           &      & 0.23     & N \\
150120A            & 0.46           & 1.2           & 1.8     & 1.4     & N \\
150423A            & 1.394           & 0.22           & 0.9     & 0.63     & N \\
150424A            &            &            & 12     & 15     & Y \\

\enddata
\end{deluxetable}

\end{document}